\documentclass[runningheads]{llncs}
\usepackage[T1]{fontenc}
\usepackage{graphicx}
\usepackage{amsmath,amssymb,url} 
\usepackage{multirow} 

\usepackage{todonotes}

\begin{document}
\title{The Effect of Perceptual Metrics on Music Representation Learning for Genre Classification}
\titlerunning{Perceptual Music Representation Learning}
%
\author{
  Tashi Namgyal$^{1}$ \quad Alexander Hepburn$^{1}$ \quad Raul Santos-Rodriguez$^{1}$\\
  \textbf{Valero Laparra}$^{2}$  \quad \textbf{Jesus Malo}$^{2}$
  \\
  $^1$University of Bristol \quad $^2$Universitat de Valencia\\
  \texttt{\{tashi.namgyal, alex.hepburn, enrsr\}@bristol.ac.uk}\\
  \texttt{\{valero.laparra, jesus.malo\}@uv.es}}

\author{Tashi Namgyal\inst{1} \and
Alexander Hepburn\inst{1} \and
Raul Santos-Rodriguez\inst{1} \and
Valero Laparra\inst{2}\and
Jesus Malo\inst{2}}
\authorrunning{T. Namgyal et al.}
%
\institute{University of Bristol, UK
\email{\{tashi.namgyal, alex.hepburn, enrsr\}@bristol.ac.uk} \and
Universitat de Valencia, Spain
\email{\{valero.laparra, jesus.malo\}@uv.es}}

\maketitle              
\begin{abstract}
The subjective quality of natural signals can be approximated with objective \emph{perceptual} metrics. Designed to approximate the perceptual behaviour of human observers, perceptual metrics often reflect structures found in natural signals and neurological pathways. Models trained with perceptual metrics as loss functions can capture perceptually meaningful features from the structures held within these metrics. We demonstrate that using features extracted from autoencoders trained with perceptual losses can improve performance on music understanding tasks, i.e. genre classification, over using these metrics directly as distances when learning a classifier. This result suggests improved generalisation to novel signals when using perceptual metrics as loss functions for representation learning.

\end{abstract}
\section{Introduction}\label{sec:introduction}

Deep learning models for music generation and understanding are increasingly being used in deployed systems. Mistakes and unnatural-sounding distortions are unacceptable to consumers and so the process of reliably evaluating the quality of model outputs is important. The gold standard for measuring the quality of generative model outputs is to gather judgements on the perceived quality from a group of human participants. However, this is a slow and expensive process, which would ideally be replaced with objective metrics that correlate well with subjective human judgements. Unfortunately, traditional training objectives such as mean squared error do not correlate strongly with such ratings \cite{vinay2022evaluating}. Instead, metrics that better match human quality ratings can be designed, \emph{perceptual metrics}, by taking into account structural information common across natural signals and/or the structure of neurological processes present in human perception. For example, in images, Structural Similarity (SSIM) \cite{wang2004image} captures structural information in natural scenes and the Normalized Laplacian Pyramid Distance (NLPD) \cite{laparra2016perceptual} mimics characteristics of the visual pathway. In addition to using these metrics for evaluation, they can act as loss functions during training. This has been shown to improve the reconstruction of images and audio when data is sparse \cite{hepburn2022on,namgyal2023data} and here we show that training models in this way can also improve performance on music understanding tasks, as demonstrated with genre classification. 

Applying and tailoring the SSIM and NLPD image quality metrics to spectrograms has recently been shown to have a better correlation with human ratings of audio quality than equivalent audio quality metrics \cite{namgyal2023what}. Despite being optimised and designed only for aligning with human perceptual opinion, perceptual image metrics have also been shown to capture statistical information about natural images~\cite{hepburn2022on,malo2010psychophysically}. This is in agreement with the \emph{efficient coding hypothesis}~\cite{barlow}, which suggests that our sensory systems have adapted to minimise redundant information when processing stimuli. This can have an advantage when training machine learning models. For instance, perceptual metrics have been used as regularisation in image-to-image translation where stochastic gradient descent leads to inaccurate gradient estimations, which the metrics protect against by including information about the distribution of natural images in the loss function~\cite{hepburn2020enforcing}. 

In \cite{namgyal2023data} we compare the reconstruction ability of compressive autoencoders trained with different metrics as loss functions, using uniform noise as input data, following~\cite{hepburn2022on}. We show that models optimised with perceptual metrics are better able to reconstruct natural audio signals at inference, despite never having seen audio signals during training. Here, we propose the use of perceptual metrics for music understanding in the setting of genre classification. Using K-Nearest Neighbours, we show a reduction in weighted F1 score when using a perceptual metric as the distance between neighbours. In contrast, we observe improved performance for a Logistic Regression classifier using latent features from autoencoders trained with perceptual metrics as the reconstruction loss.

\section {Perceptual Metrics}\label{sec:metrics}
We present two perceptual image metrics, MS-SSIM and NLPD. These are examples of a reference-based paradigm where a degraded signal is compared to a high quality reference, and their distance (or similarity) can be used to measure the perceived quality of the degraded version. While MS-SSIM and NLPD were originally designed to evaluate differences between images, it has been shown that they can be used to evaluate differences in spectrograms obtaining good agreement with human ratings of audio quality \cite{namgyal2023what}. 

\emph{Structural Similarity (SSIM)} calculates statistical features, namely luminance, contrast and structure, between local image patches. The mean and variance of these features are compared between two images to give an overall similarity score. SSIM can be calculated across multiple resolutions to calculate the Multi-Scale Structural Similarity (MS-SSIM) \cite{wang2003multiscale}, which has a stronger correlation with perceptual judgements from humans.

\emph{Normalized Laplacian Pyramid Distance (NLPD)} is a neurologically-inspired distance based on two processes found in the early stages of the visual and auditory pathways, namely linear filtering and local normalisation ~\cite{schwartz2001natural,willmore2023adaptation}. These processes reduce the redundancy found in natural signals in agreement with the efficient coding hypothesis \cite{malo2010psychophysically}. NLPD models these processes with a Laplacian Pyramid \cite{burt1983laplacian}, followed by a divisive normalization step at each stage of the pyramid \cite{malo2010psychophysically}. The resulting distance measure correlates well with human perception of images \cite{laparra2016perceptual} and audio \cite{namgyal2023what}.

The structures that these perceptual metrics pick up on can be visually compared using a tool such as IQM-Vis~\cite{iqmvis}, which highlights sensitivities or invariances to certain transformations such as change in brightness, rotation or contrast.

\section{Genre Classification}

In this section, we consider the relevance of perceptual features to music understanding, specifically to a genre classification task. We use the GTZAN dataset \cite{tzanetakis_essl_cook_2001}, which contains 1000 samples across 10 genres, with 70 duplicated or distorted signals removed as indicated in \cite{Sturm_2014}. This results in unequal class sizes so we use a weighted F1 score to evaluate class predictions (Table \ref{tab:classes}). 

\begin{table}[!htb]
    \centering
    \caption{Class sizes in the filtered GTZAN dataset, reducing the dataset from 1000 to 930 songs.}
    \begin{tabular}{|c|c|c|c|c|}
        \hline
        Genre & Total & Train & Valid & Test \\
        \hline\
        Blues & 100 & 46 & 23 & 31\\
        Classical & 99 & 48 & 20 & 31\\
        Country & 98 & 45 & 23 & 30\\
        Disco & 93 & 42 & 22 & 29 \\
        Hip Hop & 92 & 47 & 18 & 27\\
        Jazz & 87 & 43 & 17 & 27\\
        Metal & 91 & 44 & 20 & 27\\
        Pop & 84 & 41 & 13 & 30\\
        Reggae & 86 & 43 & 17 & 26\\
        Rock & 100 & 44 & 24 & 32\\
        \hline
    \end{tabular}
    
    \label{tab:classes}
\end{table}

\begin{figure}[!htbp]
    \centering
    \includegraphics[width=\columnwidth]{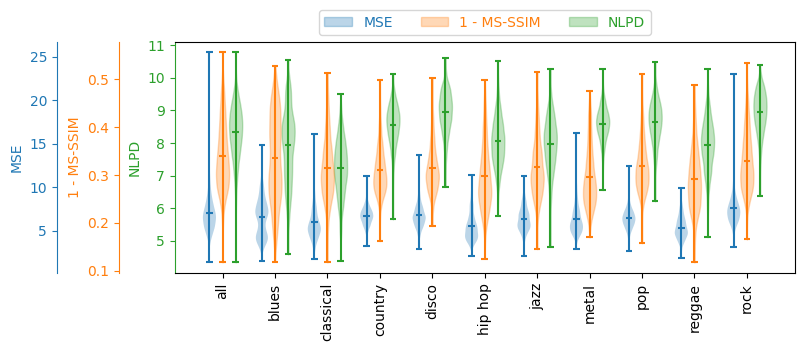}
    \caption{Violin Plot showing the distribution of pairwise distances between songs for each genre in the filtered GTZAN dataset. MSE, $1-\text{MS-SSIM}$ \& NLPD have different distributions. The `all' column shows pairwise distances across the whole dataset and each genre column shows pairwise differences within that genre. Within genre distributions are more spread out for NLPD than MSE. This effect is detrimental to clustering, but advantageous to reconstruction, where small perceptual differences need to be highlighted to increase their impact on the loss for improved learning.}
    \label{fig:violin}
\end{figure}

First we compare the usefulness of perceptual metrics (MS-SSIM and NLPD) with MSE in a clustering approach. MS-SSIM is a similarity metric in the range [0,1], so we convert it to a distance by subtracting from 1. We calculate MSE, $1-\text{MS-SSIM}$ and NLPD between the precomputed mel-spectrograms provided as part of the GTZAN dataset, with the zero-padding removed. We calculate the pairwise distance between each song across the entire dataset, then calculate the pairwise distances for each song within each genre. We show that the distribution of pairwise distances varies considerably between the metrics (Fig. \ref{fig:violin}).

\begin{figure}[!htbp]
    \centering
    \includegraphics[width=\columnwidth]{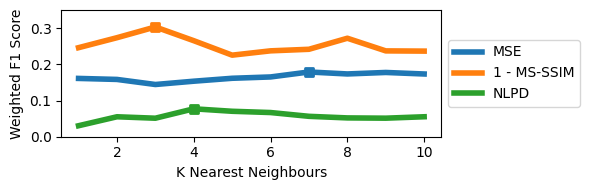}
    \caption{Weighted F1 score on the validation set for KNN classifiers using MSE, $1-\text{MS-SSIM}$ \& NLPD as distances between neighbours. Squares show the number of neighbours, $k$, chosen for each model.}
    \label{fig:f1}
\end{figure}

Additionally, we test the ability of the distances to cluster data into genres. We use a K-Nearest Neighbours classifier, which uses the pairwise distance between samples. The number of neighbours for each metric in selected based on a weighted F1 score with the training and validation splits. We use 7, 3 and 4 neighbours for MSE, $1-\text{MS-SSIM}$ and NLPD respectively (Fig. \ref{fig:f1}).

We retrain 3 compressive convolutional autoencoders from \cite{namgyal2023data} with uniform noise, using MSE, $1- \text{MS-SSIM}$ and NLPD as loss functions. We pass the original padded spectrograms through the trained network to extract the compressed latent representations for each spectrogram. These features are then used as inputs to a logistic regression classifier, the idea being that the compressed representation when optimised for a perceptual distance will contain features useful to determining the genre. We set a maximum entropy of the embedding, by projecting values in the embedding to predetermined set of integers. This is to ensure a fair comparison of features learnt by using the different loss functions.

\subsection{Results}
\begin{figure}[!htbp]
    \centering
    \includegraphics[width=1\columnwidth]{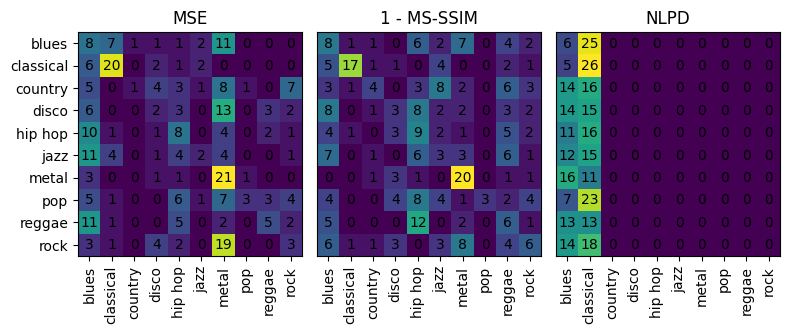}
    \caption{Confusion Matrices for KNN classifiers using MSE, NLPD and 1 - MS-SSIM as distances between neighbours.}
    \label{fig:cm}
\end{figure}

The results for the K-Nearest Neighbours classifier indicate that MSE and $1-\text{MS-SSIM}$ are more appropriate than NLPD (Table \ref{tab:mucresults}) for measuring the distance between samples. We also demonstrate this with confusion matrices (Fig. \ref{fig:cm}). The NLPD model heavily overpredicts the blues and classical genres while the MSE and $1-\text{MS-SSIM}$ models have more balanced predictions. We see improved results using the latent features obtained from the autoencoders (Table \ref{tab:mucresults}). 

\begin{figure}[!htbp]
    \centering
\includegraphics[width=1\columnwidth]{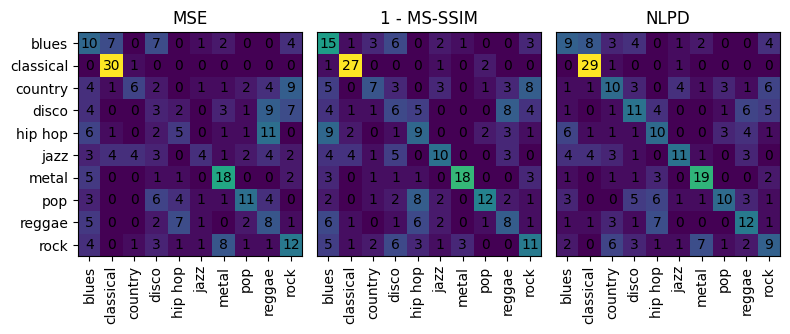}
    \caption{Confusion Matrices for Logistic Regression classifiers using latent features from autoencoders trained on uniform noise with MSE, NLPD and 1 - MS-SSIM distances as losses.}
    \label{fig:cm2}
\end{figure}

The metrics perform worse as distances in the KNN classifiers since they rely on low level features of the signal, while the autoencoders trained with them can extract higher level features that may be more directly related to genre. The logistic regression classifiers trained with NLPD and $1-\text{MS-SSIM}$ give higher F1 scores than the model trained with MSE when using the autoencoder features. The classical genre is sufficiently different from the other classes to be predicted well in all cases but the perceptual models give more balanced predictions across the remaining classes (Fig. \ref{fig:cm2}). 
$1-\text{MS-SSIM}$ performs relatively well in both experiments. However, these results suggest that NLPD is more appropriate for use as a reconstruction loss where the input and output are very similar, but not for clustering distances where samples can differ significantly. This may be because NLPD acts to remove redundant information in the data, which would remove slow-changing parts of signals that are more likely to intersect and emphasise any fast-changing parts of the signal that are less likely to co-occur. We see this in Fig. \ref{fig:violin} where within genre distributions are more spread out for NLPD than MSE. This effect is detrimental to clustering, but advantageous to reconstruction, where small perceptual differences need to be highlighted to increase their impact on the loss for improved learning. It is worth noting that our models are relatively simple and both sets of results are far from the state-of-the-art for the GTZAN dataset. Several papers previous to 2014 reported accuracy above 95\%  \cite{Sturm_2014}, though these results were called into question due to duplicated and distorted data. More recently, \cite{chen_audio_2024} claimed their F1 score of 92.47\% using reservoir computing was better than other contemporary state-of-the-art systems, though this paper also used the full dataset. These other models use different input features and architectures of varying complexity and retraining compatible models with perceptual losses is left to future work.

\begin{table}[!htb]
    \centering
    \caption{1: Weighted F1 Score for KNN classifier using pairwise MSE, NLPD and $1-\text{MS-SSIM}$ distances. 2: Weighted F1 Score for Logistic Regression classifier using latent features from autoencoders trained with MSE, NLPD and 1 - MS-SSIM losses.}
    \begin{tabular}{|c|c|c|}
        \hline
        
        Experiment & Metric & Weighted F1 Score \\
        \hline
        \multirow{3}{2.5cm}{\centering KNN on Pairwise Distances} & MSE & 0.220 \\
                                                   & 1 - MS-SSIM & \textbf{0.266} \\
                                                   & NLPD & 0.035 \\
        \hline
        \multirow{3}{2.5cm}{\centering LR on Latent Features} & MSE & 0.355  \\
                                               & 1 - MS-SSIM & 0.426 \\
                                               & NLPD & \textbf{0.439} \\
        \hline
    \end{tabular}

    \label{tab:mucresults}
\end{table}

\section{Conclusion}\label{sec:conclusion}

Perceptual metrics that are used across different modalities and domains are assumed to capture information about the structure of natural signals. As shown previously, perceptual metrics can be used in lieu of natural data and still be able to reconstruct natural signals at test time~\cite{hepburn2022on,namgyal2023data}. Here, we also show that this can lead to improved representations when performing genre classification. We show that perceptual metrics transform signals in different ways and so care should be taken over the choice of metric depending on the task. These perceptual metrics are somewhat sensitive to spectrogram and filter parameter settings, though more reference and distorted audio with corresponding human quality labels is needed to reliably quantify the extent. Nevertheless, these results pave the way for principled training and evaluation of generative audio and representation models with perceptual metrics, while potentially alleviating the requirement for prohibitively large-scale datasets. 

\begin{credits}
\subsubsection{\ackname} TN is supported by the UKRI AI CDT (EP/S022937/1). AH and RSR are supported by UKRI Turing AI Fellowship EP/V024817/1. VL and JM are supported by MINCEO and ERDF grants PID2020-118071GB-I00, DPI2017-89867-C2-2-R and GV/2021/074.
\subsubsection{\discintname}
The authors have no competing interests to declare that are
relevant to the content of this article. 
\end{credits}

%
%
%
\bibliographystyle{splncs04}
\bibliography{refs}
%




\end{document}